**Extragalactic Background Light: Inventory of light throughout the cosmic history**

Kalevi Mattila (1) and Petri Väisänen (2,3)

  (1) Department of Physics, University of Helsinki, Finland; e-mail: kalevi.mattila@helsinki.fi
  (2) South African Astronomical Observatory, Cape Town, South Africa;
  (3) Southern African Large Telescope, Cape Town, South Africa

ABSTRACT
The Extragalactic Background Light (EBL) stands for the mean surface brightness of the sky as we would see it from a representative vantage point in the intergalactic space outside of our Milky Way Galaxy. Averaged over the whole $4\pi$ solid angle it represents the collective light from all luminous matter radiated throughout the cosmic history. Part of the EBL is resolved into galaxies that, with the increasing detecting power of giant telescopes and sensitive detectors, are seen to deeper and deeper limiting magnitudes. This resolved part is now known to contribute a substantial or even the major part of the EBL. There still remains, however, the challenge of finding out to what extent galaxies too faint or too diffuse to be discerned individually, individual stars or emission by gas outside the galaxies, or – more speculatively – some hitherto unknown light sources such as decaying elementary particles are accounting for the remaining EBL. We review the recent progress that has been made in the measurement of EBL. The current *photometric* results suggest that there is, beyond the resolved galaxies, an EBL component that cannot be explained by diffuse galaxy halos or intergalactic stars.

# 1. Introduction

## 1.1 Why is the sky dark at night: Olbers' Paradox

The idea of an unbound (and static) Universe populated with stars like our Sun entered into the minds of astronomers and other scientists and philosophers already in the sixteenth and seventeenth century. As consequence of this the question "why then is the sky dark at night" was asked by Thomas Digges, Johannes Kepler, Otto von Guericke, Christian Huygens and Edmund Halley among the first [1–3]. A precise formulation of the problem and a suggestion for its solution were for the first time presented by Jean-Philippe Loys de Chéseaux [4] in 1744: if the Universe is infinite in size and homogeneously populated by stars then visible disks of the stars should cover the whole sky. Why then is the sky not everywhere as bright as the surface of the Sun? His solution was that an obscuring medium hides the more distant stars from our view. Heinrich Olbers, independently, presented in 1826 the same problem and a similar solution in terms of interstellar absorption [5]. However, the absorption of visible light would lead to heating of the obscuring matter, and the thermal re-radiation of the absorbed energy would only move the riddle to infrared and longer wavelengths.

   More than a century later, in the context of relativistic expanding models of the Universe the dark-sky riddle experienced a reincarnation. In the case of an infinite Universe uniformly filled with galaxies (instead of stars) one encounters the same riddle again: by considering just the extragalactic component of the background light (EBL) one would again end up with a sky that is uniformly covered by disks of the stars in galaxies. In his classical text Cosmology Bondi [6], referring to Olbers' 1826 paper, coined the riddle the name 'Olbers' Paradox' and offered the redshift as a solution to it. However, Harrison [7] first showed that in an evolving Universe of finite age the redshift has only a minor contribution to the solution. The main effect comes from the finite age of the stars, a solution offered in the long history of the paradox already by Kelvin (see [8]).

   In the past decades the importance of the extragalactic background light for cosmology has been repeatedly emphasized, see e.g. [9] and for a review [10]. Some central, but still largely open astrophysical problems to which EBL measurements can shed new light include the formation and early evolution of galaxies and the star formation history of the Universe. A large fraction of the energy released in the Universe since the epoch when the first galaxies formed is expected to be contained in the EBL.

The EBL at optical, ultraviolet and near infrared wavelengths consists of the integrated light of all galaxies along the line of sight plus any contributions by individual stars or emission by gas outside the galaxies; also - more speculatively - some hitherto unknown light sources such as decaying elementary particles could contribute to it. There may also be a significant number of galaxies that, due to their extreme faintness or low surface brightness, escape detection as discrete objects but contribute to the cumulative EBL [11,12].

In observational cosmology measurements of the background brightness have, in principle, an advantage over the number count observations. When counting galaxies, whether in magnitude or redshift bins, one needs to consider many kinds of selection effects which affect the completeness of the sample. The measurement of the EBL is not plagued by this particular problem. However, it is hampered by the foreground components, much larger than the EBL itself, and their elimination or accurate evaluation is of key importance for the direct photometric measurement of the EBL.

## 1.2 View of background radiations over the whole electromagnetic spectrum

The discovery of the 3 K Cosmic Microwave Background (CMB) in 1965 by Penzias and Wilson [13] meant the confirmation of the evolving big-bang model and created a new CMB–dominated era in observational cosmology. It inspired work also at other wavelengths. Thus, the radiation backgrounds from $\gamma$– and X–rays over ultraviolet, optical, near– and far–infrared to long radio waves were recognized as important sources of information for cosmology and galaxy formation studies and they have been intensively studied both by ground–based and spaceborne telescopes. Energetically, the optical–near-IR and the far-IR components are – after the CMB – the next most important ones (see Fig. 1). The source of energy for both of them is thought to be mainly the starlight, the direct starlight in optical and near-IR and the dust–processed one in the mid- and far-IR. Besides the nucleosynthesis in stars, also accretion to active galactic nuclei plays a substantial role as energy source [14].

An important aspect is the balance between the ultraviolet–to–near-IR ($\lambda \approx 0.1 - 3$ μm) and the mid–to–far-IR ($\lambda \approx 5 - 300$ μm) EBL: what is lost through absorption by dust in the ultraviolet-to-near-IR will re-appear as emission in the mid–to–far-IR. A dusty galaxy spectrum has two bumps, one in the optical-near-IR, the other in the far-IR. This aspect is strongly emphasized by the detection of the far-IR EBL [15–17] at an energy level comparable to that of the optical EBL estimates.

Reviews of cosmic background radiation over the whole electromagnetic spectrum have been presented by e.g. in [10,18–22]. Several conferences have been devoted to the different backgrounds and their interconnections, see [23–27]. For reviews concentrating especially on the optical-infrared EBL we refer to [28–33].

Like for the CMB fluctuations also the anisotropy of the backgrounds at other wavelengths, from X-rays to far-IR, has become subject of extensive studies, for a review see e.g. [34]. In the present review we will concentrate mainly on the optical mean EBL only.

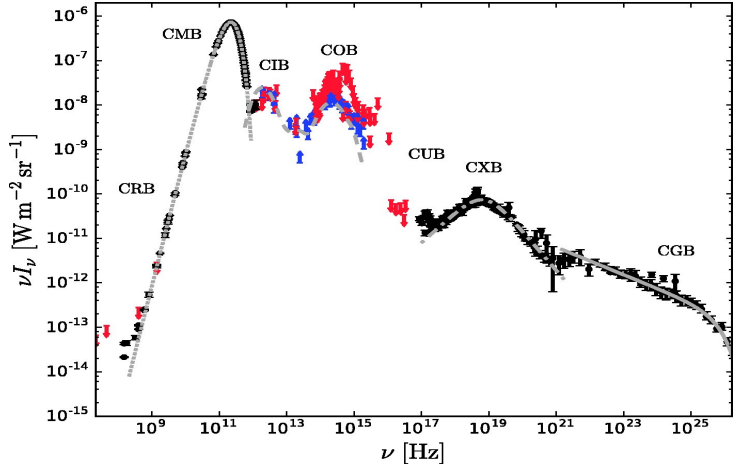

Figure 1. Complete cosmic background radiation, or extragalactic background light, from the radio (CRB) and the well-known micro-wave background (CMB), to X-rays (CXB) and γ-rays (CGB) at the high energy end. This review focuses on the optical background radiation (COB in the figure), as well as the neighbouring IR and UV regimes (CIB, CUB). Black points with error bars indicate detections, while blue and red arrows indicate lower and upper limits, respectively. The gray lines are models of various sections of the cosmic background (adopted from Hill et al. [22], Fig. 8, courtesy of Applied Spectroscopy).

## 2. Photometric/spectrophotometric measurement of the optical EBL

Contrary to the expectations for an infinite static Universe, the night sky is not blazingly bright; however, one might still have expected that adding up the contributions from all galaxies and other luminous matter throughout the Universe would result in an easily observable optical night sky brightness. This is, however, not the case and, paradoxically, the optical EBL has remained the least-well-known component after the microwave, X-ray, γ-ray, infrared, and radio backgrounds have been detected and reasonable well characterized.

In this section we will discuss the direct measurements of the optical EBL. In the next two sections a brief review is given on the recent measurements in the neighbouring UV and near-IR wavelength regions, and on the method of using γ-ray attenuation mesurements for the determination of the EBL. In Section 5 we will then discuss the EBL produced by the integrated light of galaxies and whether it can alone account for the EBL.

### 2.1 Components of the Light of the Night Sky

The conventional approach to EBL measurement has been to observe the total sky brightness and to measure or model all the foreground components. After foreground subtracion, what is left over is the EBL. In this method, because of smallness of the EBL, the total sky brightness and each foreground components must be very accurately known. In addition, if measured by different telescopes or methods, they must be absolutely calibrated at a level of ≤ 0.5% which, for surface brightness measurements, is hardly possible [35].

When all or part of the observations are done from ground the first hindrance is the atmospheric diffuse light, consisting of the airglow and the tropospheric scattered light. Airglow is created mainly in the ionospheric E layer at ~90 km and F layer at ~150 km as result of the recombination of excited states of common atmospheric gases.

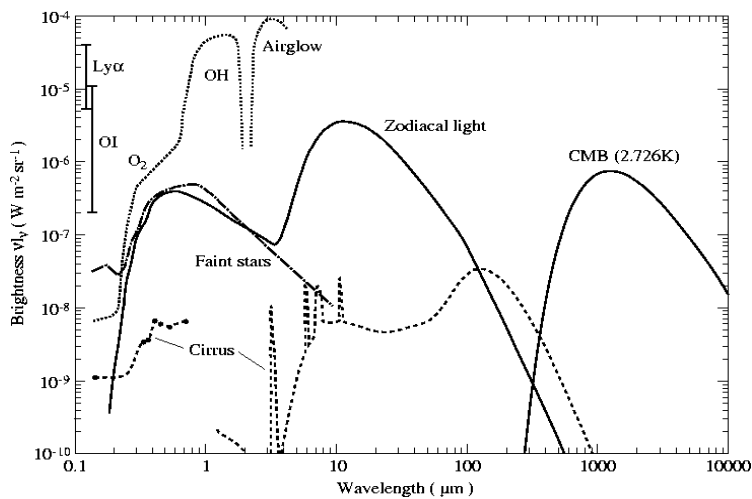

Figure 2. Overview of the night sky brightness as seen outside of the lower Earth's atmosphere and at high galactic and ecliptic latitudes. The zodiacal emission, zodiacal light, and starlight (with bright stars excluded) are for the South Ecliptic Pole. The interstellar cirrus component is normalized for a column density of 1020 H-atoms/cm2 which is close to the values at the darkest patches in the sky. For the Airglow the labels indicate important agents in the optical (O2 ) and near-IR (OH). The geocoronal Lyman α (121.6 nm) and O I (130.4, 135.6 nm) line intensities are as measured with the Faint Object Camera of the Hubble Space Telescope at a height of 610 km. (adopted from Leinert et al. [36] wherein the references and more detailed explanations are given.) Credit: C. Leinert et al., A&AS, 127, 2, 1998, reproduced with permission c ESO.

**Table 1.** Typical values of the sky background brightness components in the blue (B) spectral region: Zodiacal Light (ZL), atmospheric diffuse light (ADL) which is the sum of airglow (AGL) and tropospheric scattered light, integrated starlight (ISL), diffuse galactic light (DGL) and the integrated light of galaxies (IGL). All values refer to outside the atmosphere, except for $I_{ADL}$ which is the value below the troposphere, at ground-level.

| Component | Wavelength [nm] | Value [nW m$^{-2}$sr$^{-1}$] | Comment/reference |
|---|---|---|---|
| $I_{ADL}$ | $400 - 420$ | $330 - 570$ | near Zenith, [36] |
| $I_{ZL}$ $^a$ | $400 - 420$ | $370 - 410$ | for $|\beta| \geq 35^o$, [36] |
| $I_{ISL}$ | $440$ | $150 - 290$ | $|b| \geq 35^o$, for stars with $m \geq 6.5^m$, [37] |
| $I_{DGL}$ | $440$ | $30 - 60$ | $|b| \geq 35^o$, [36,37] |
| $I_{IGL}$ | $470$ | $5.3 \pm 0.9$ | [38], including all galaxies |

$a$ derived from $I_{ZL}$ at 500 nm using the Solar spectrum of [39] and ZL colour of Leinert et al. [36]

The tropospheric layers of gas and aerosols at < 10 km height create a sky brightness component via (multiple) scattering of the combined light of airglow and all the other night sky components in analogy to the blue daylight sky. The second large foreground is the zodiacal light, which is created by scattered sunlight off the interplanetary dust particles. At heights  >~150 km the atmospheric components are left below and the zodiacal light dominates the diffuse sky brightness. A very promising approach is to make the measurements beyond the interplanetary dust cloud. However, unless an imaging detector or a small and accurately positioned photometer aperture can be used, the integrated light from stars enters into the measured signal and over-whelms the EBL. On the other hand, using large ground-based telescopes the starlight can be excluded to faint limiting magnitudes. Even if all these foregrounds have been successfully determined there remains the emission by interstellar gas and the diffuse galactic light, the starlight scattered by interstellar dust particles, well known via the attenuation they cause to starlight (interstellar extinction). No perfect 'cosmological windows', completely free of Galactic foreground gas and dust, are known to exist.

Typical values of the different sky brightness components in the optical spectral region are given in Table 1. **A note on the units:** In this article the surface brightness is expressed mostly as $\lambda I_\lambda = \nu I_\nu$ and is given in units of nW m$^{-2}$ sr$^{-1}$. In some cases we give the observed surface brightness $I_\lambda$ also in units of $10^{-9}$ erg cm$^{-2}$ s$^{-1}$ sr$^{-1}$ Å$^{-1}$ In Astronomy the magnitude system is commonly used. As measure of the flux $F_\lambda$ of a star or galaxy one has its magnitude m = $-2.5$log $F_\lambda$ + C and, instead of the surface brightness $I_\lambda$, correspondingly $\mu = -2.5$log $I_\lambda$ + C in units of mag/arcsec$^2$. The magnitudes increase from m = -1.6 to +6 for naked eye stars to m $\sim$ 30 for the faintest objects currently discernible with big telescopes.

## 2.2 Separation methods and early measurements of the EBL

The situation in the optical is very different from that for the microwave band where the extragalactic component (2.7 K) is roughly equal to the atmospheric one ($\sim$2.3 K; [13]), and both the solar system and Galactic components are much smaller.

The faintness of the EBL is not the main problem. Discrete sources, e.g. stars or galaxies, are today observed and even spectra are obtained with large telescopes at the very low flux levels of 28 mag or even fainter. The optical EBL intensity (surface brightness) is $\sim$28 mag arcsec$^{-2}$. Thus the EBL signal from a $10 \times 10$ arcsec$^2$ spot on the sky corresponds to a discrete object of $\sim$23 mag, a rather easy task for today's large telescopes. The problem is, however, that nowhere in the sky can one make an easy differentiation between an EBL-on and EBL-off position as is the case for small-size discrete objects like distant galaxies.

A basic difficulty when trying to measure the EBL is its isotropy: its intensity on large scales is expected to be (almost) constant all over the sky. Furthermore, the EBL spectrum is expected to be a smooth one without any characteristic spectral features. This can be understood because the radiation from galaxies and other luminous matter over a vast redshift range contributes to the EBL, thus washing out any spectral lines or discontinuities.

Three basic approaches have been used in the measurements of the EBL:
(1) utilize the difference of the spatially variable distribution of the foreground versus the isotropic distribution of the EBL intensity; presuming that the zero point for the foreground is known by some other method (e.g. infrared emission, hydrogen 21-cm line) one can estimate the EBL by extrapolation to this zero point;
(2) utilize the difference between the spectra of the EBL (featureless) and the foreground components having spectral features; for the surface brightness component separation an early example is the analysis of the Solar corona for which the dust-scattered F corona was detected and separated from the continuous-spectrum electron-scattering K corona by means of its Fraunhofer absorption line spectrum [40];
(3) because both these methods have difficulties, one should try to eliminate or minimize as many as possible of the foreground components by a suitable selection of the observing site and of the observing technique; e.g. by making the observations from outside the Earth's atmosphere or even from the outer Solar System, beyond the interplanetary dust cloud.

The first attempt to measure the EBL photometrically by Roach and Smith [41] was based on the idea that, as seen from our vantage point inside the Galaxy, the observed intensity of the EBL is not constant but decreases towards the galactic plane in accordance with the increasing extinction through the galactic dust layer. This fore-ground extinction also causes the 'zone of avoidance' phenomenon: less galaxies are seen towards the galactic plane than towards the pole regions [42]. However, the two major Galactic light components, integrated starlight and diffuse galactic light, also show a dependence on the galactic latitude and completely overwhelmed the EBL.

## 2.3 EBL measurements from space and ground

A unique opportunity for EBL measurement was opened up by the Pioneer 10 and 11 spacecraft (Fig. 3) as they in 1972 –1974 passed outside the asteroid belt to > 3 astronomical units and the zodiacal light dropped to vanishingly small values. The only remaining sky components were the integrated starlight, diffuse galactic light and EBL. Because of the large field of view, $2.3 \times 2.3$ deg$^2$, of the Pioneer 10/11 Imaging Phopolarimeter Instrument, starlight unavoidably dominated the signal and a large starlight correction was needed. The diffuse galactic light contribution was estimated from modelling. An upper limit to the EBL at 440 nm was presented by Toller [43].

Matsuoka et al. [44] re-anayzed the Pioneer 10/11 data and announced a detection of the EBL at 1.5- to 2-$\sigma$ level: $\lambda I_{EBL}$ = 7.9±4.0 and 7.7±5.8 nW m$^{-2}$ sr$^{-1}$ at 440 nm and 640 nm, respectively. Although virtually free of the zodiacal light contamination, this measurement is plagued by the problem that the EBL is derived as the small difference between two much larger quantities, the total sky brightness in the field of view of the instrument and the integrated light of all stars fainter than 6.5 mag which could not be individually eliminated. The background sky brightness as seen in the field of view, $I_{total} > \sim 180$ nW m$^{-2}$ sr$^{-1}$, consisted to $> \sim 95\%$ of the integrated starlight and diffuse galactic light. To derive the integrated starlight intensity a number of diverse star catalogues, which had different photometric systems, had to be compiled. For a detection of the EBL a very high systematic accuracy ( $< \sim 1\%$) of the absolute photometric calibrations, separately for the Pioneer 10/11 surface photometry and for the catalogue-based estimates of the integrated starlight would have been required. Considering the several systematic error sources involved in the integrated starlight and diffuse galactic light estimation, it is not clear whether this goal has been achieved. Recently, a renewed examination [45] of the dataset [44] has revealed significant, occasionally occurring instrumental offsets as well as extra noise due to external signals of unknown origin. Because of this, the authors [45] consider the data set to be qualified only for relative sky brightness measurements but not for purposes requiring absolute sky brightness values as was the case for the determination of the mean EBL intensity in [44].

Recently, the Long Range Reconnaissance Imager instrument aboard NASA's New Horizons Pluto mission (see Fig. 3) acquired broad band (440 – 870 nm) sky background measurements of four suitable areas during its cruise phase beyond Jupiter's orbit, at 7 to 17 astronomical units [46]. The sky background in the anti-solar hemisphere was virtually free of zodiacal light and, thanks to the good spatial resolution (1" $\times$1" ) of the instrument, sky background intensities could be determined with much less starlight contamination than in the Pioneer 10/11 photometry. The diffuse sky background values as observed in the available four fields are shown in Fig. 4. After masking of stars brighter than 17.5 mag the only remaining components are the diffuse galactic light, the residual light from faint (m > 17.5 mag) stars, and the EBL. One can see that the observed total sky background reaches a very low level of $\sim 30 - 50$ nW m$^{-2}$ sr$^{-1}$ which is only $\sim 2 - 4$ per cent of the sky brightness at good dark observatory sites on remote mountain tops [47,48]. These data have been used by Zemcov et al. [46] to derive an upper limit to the EBL intensity. The limiting factor was the relatively large contribution by the diffuse galactic light. This could be estimated only via modelling; this included substantial uncertainties concerning both the dust column density estimates via 100µm emission, as well as the scattering properties of the interstellar dust grains. A 2-$\sigma$ statistical upper limit to EBL was set at $I_{EBL} < 19.3$ nW m$^{-2}$ sr$^{-1}$. Including their estimated systematic errors of +10.3/-11.6 nW m$^{-2}$ sr$^{-1}$, mainly resulting from the diffuse galactic light model estimates, this leads to an EBL overall upper limit of 29.6 nW m$^{-2}$ sr$^{-1}$ (see Table 2 and Figs. 4 and 8).

A spectrum-based measurement of the EBL was presented by Dube et al. [50, 51]. In the pre-CCD era a photoelectric photometer was used. The stars were blocked out with specific masks. The zodiacal light was separated on the basis of its solar-like spectrum which displays the strong Mg I Fraunhofer line doublet at 5175 Å that, of course, is not present in the EBL. The airglow and the tropospheric scattered light were measured as function of the zenith distance and, assuming that they are proportional to the layer thickness through the atmosphere, sec z, were extrapolated to outside the atmosphere, to sec z = 0. A basic problem was that the total brightness of the night sky was ca. 100 times the EBL. Furthermore, no attempt to separate the diffuse galactic light was done.

The spectroscopic method for separation of the zodiacal light was used again by Bernstein et al. [52, 53, 54] who announced that they had achieved 'the first detection' of the EBL at 300, 550, and 800 nm. A combination of space-borne (Hubble Space Telescope) and ground based (Las Campanas Observatory) measurements was used. While the total sky brightness measured from space was free of atmospheric effects this was not the case for the equally important zodiacal light brightness; it had to be measured from the ground and amounted to >~95% of the total sky intensity. Its measurement was entangled with the full complexity of the strong atmospheric signals, the airglow and the tropospheric and ground-reflection (multiply) scattered light components. In their analysis, however, they neglected some important effects of the atmospheric scattered light. For the absolute surface photometric calibration they were not able to achieve the accuracy of <~0.5%, required for both the space borne and the ground based measurements. Therefore, the claim for a detection of the EBL appeared premature [35]. After reanalysis of their systematic errors in [55] and [56] the authors came to the same conclusion, namely that

' ... *the complexity of the corrections required to do absolute surface (spectro)photometry from the ground make it impossible to achieve 1% accuracy in the calibration of the ZL', and*
'...*the only promising strategy ... is to perform all measurements with the same instrument, so that the majority of corrections and foreground subtractions can be done in a relative sense, before the absolute calibration is applied.*'
As the final result of the five papers [52–56] we quote from [56]: *"The EBL23 results we obtain are roughly 1-2$\sigma$ detections and can be quoted as upper limits at the +2$\sigma$ values on their own."* These upper limits are given in Table 2.

## 2.4 EBL measurement using dark cloud shadow

A method that combines the elimination of the zodiacal light and airglow foreground components via differential measurements, and the separation of the diffuse galactic light using its spectral signature has been presented by Mattila [31, 57]. The screening effect of a dark cloud on the background light is being utilized. The difference of the night sky brightness in the direction of an opaque high galactic latitude dark cloud and a transparent sky area is due to two components only: (i) the EBL which passes freely along the transparent line of sight but is blocked towards the cloud and (ii) the scattered starlight from the dust in the cloud, only weakly present in the transparent sky position. The main task in the method is to separate the scattered starlight of the dark cloud itself by means of spectroscopy. While the scattered starlight spectrum has the characteristic stellar Fraunhofer lines and the discontinuity at 400 nm the EBL spectrum is a smooth one without these features. Because of the differential measurement method the three large foreground components, i.e. the zodiacal light, the airglow, and the tropospheric scattered light, are eliminated.

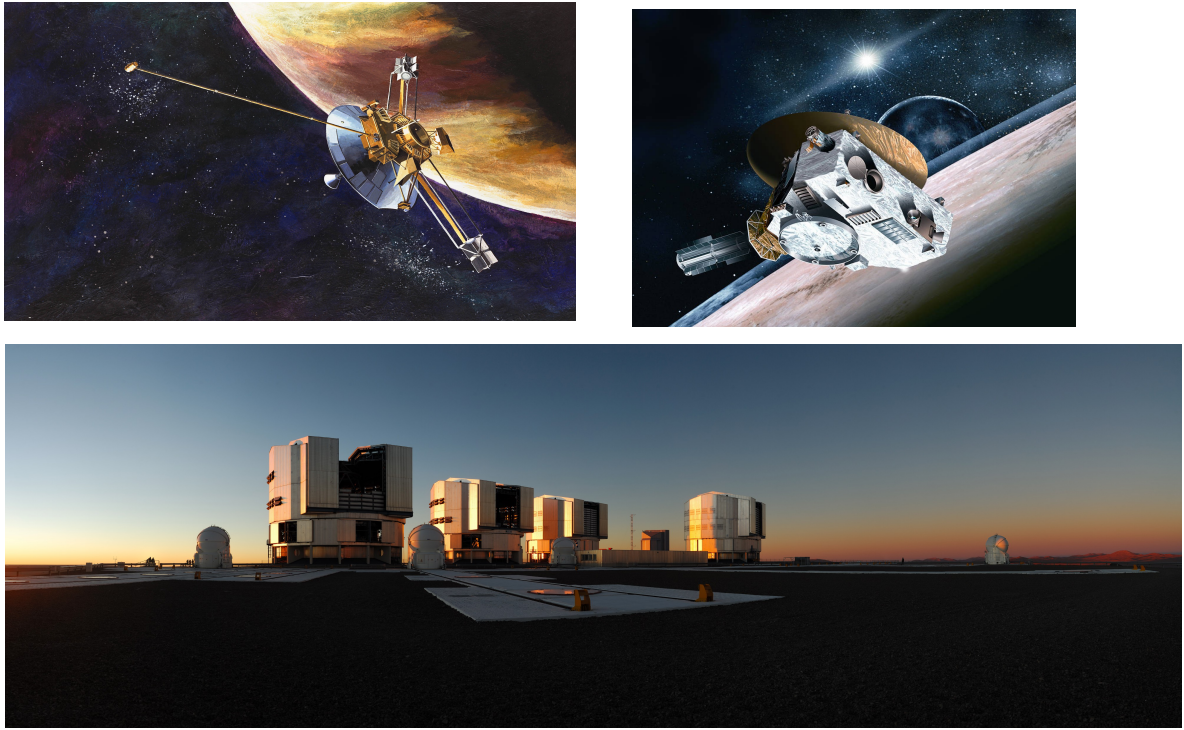

**Figure 3. I**nstruments used for optical EBL measurements. upper left: Pioneer 10; upper right: New Horizons; (Credit: The two images courtesy of NASA) bottom: European Southern Observatory's Very Large telescope at Cerro Paranal. (Credit: Image courtesy of ESO).

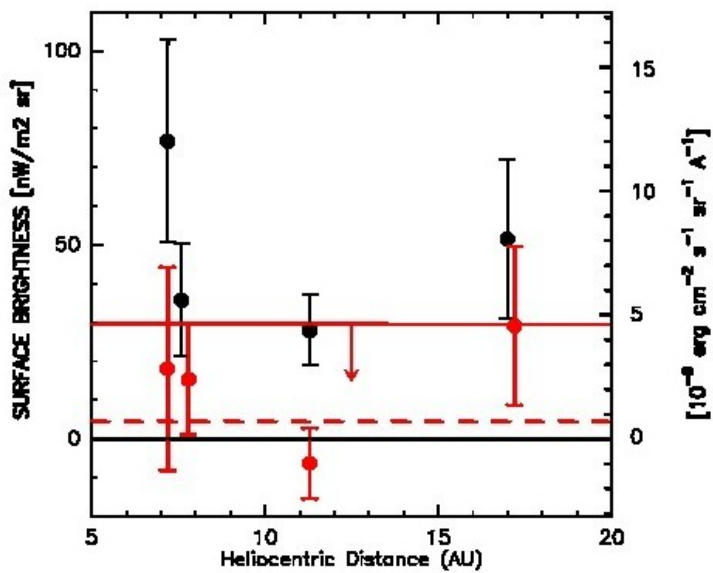

Figure 4. Measurements by the New Horizons Long Range Reconnaissance Imager instrument of the sky background brightness at $440 - 870$ nm in four fields taken beyond the interplanetary dust cloud at heliocentric distances between 7.6 and 17.0 astronomical units, Zemcov et al. [46]. Black dots indicate the total diffuse sky values with stars brighter than 17.75 mag masked out. The red points indicate the residuals together with 1-σ statistical errors after model-estimated values for faint stars and diffuse galactic light have been subtracted; the dashed red line is their mean value. The solid red line with downwards-pointing arrow indicates the EBL upper limit, including systematic and 2-σ statistical errors. For comparison the zodiacal light sky brightness at 1 astronomical unit towards the anti-solar hemisphere at high ecliptic latitudes is $\sim$500 nW m$^{-2}$ sr$^{-1}$ [36]. Passing through the asteroid belt between 2.2 - 3.5 astronomical units the zodiacal light bri ghtness drops significantly and is consistent with zero intensity beyond this distance [49].

Fig. 5a illustrates the geometry of the night sky components involved; it is a kind of geocentric 'onion-skin world model' showing the different layers of light that contribute to the night sky brightness. Three observing positions are schematically indicated in Fig. 5b: '2' within the opaque dark cloud, '1' and '3' in transparent sky areas. If the the scattered light from dust in the cloud were zero the sky brightness at the position of the opaque cloud would drop by an amount equal to the intensity of the EBL (dashed line in Fig. 5c). Because of the scattered light (shaded area in Fig. 5c) the sky is, however, brighter in the direction of the dark cloud than outside of it. Fig. 5d demonstrates the difference in the spectral shape of the Galactic starlight (Fraunhofer lines) and the 'expected' shape of the EBL (smooth spectrum).

Besides in [31,57] the dark-cloud method was used also in [58] at 400 nm and in [59] in the near-IR. In an innovative attempt the Jovian satellites during eclipse phase have been utilized, instead of a dark cloud, as opaque foreground screens [60].

Recent EBL measurements in the area of the high-latitude dark nebula Lynds 1642 have been presented by Mattila et al. [61, 62]. One and the same instrument, FORS1 or FORS2 at ESO's Very Large Telescope (see Fig. 3) was used for the differential measurements. No sky components needed to be measured or evaluated separately, with another instrumenet; therefore, unlike in the methods based on the total sky brightness, no higher-than-usual calibration accuracy was required. In Fig. 6 (upper panels) images of two observed fields in and near Lynds 1642 are shown: the opaque core of the cloud with visual extinction of >15 mag on the left and a transparent sky position on the right. The differential spectrum, $\Delta I(\lambda)$, cloud position minus transparent sky position, is shown in the bottom panel as black line. It is seen to be closely similar to the integrated starlight spectrum as shown in Fig. 5d, modified, however, by the reddening caused by extinction in the cloud. Suitable spectral features for the separation of the scattered light are the 400 nm discontinuity, the strong Fraunhofer lines H and K of Ca II at 397 and 393 nm, and the G band at 430 nm. To a lesser degree the Mg I+MgH band at 517 nm and the Fe line at 527 nm can be useful. The blue curve is the scattered light spectrum for the transparent (off) positions, and the red curve shows the Balmer emission line correction for the line-of-sight gas.

Fig. 7 demonstrates qualitatively that the spectroscopic separation method is capable of reaching the required sensitivity level of $<\sim 1\ 10^{-9}$ erg cm$^{-2}$ s$^{-1}$ sr$^{-1}$ Å$^{-1}$ or $<\sim 4$ nW m$^{-2}$ sr$^{-1}$. Model fits, superimposed on the observed spectrum, are shown for EBL intensities of 0 (blue), 3 (red), and 8 $10^{-9}$ erg cm$^{-2}$ s$^{-1}$ sr$^{-1}$ Å$^{-1}$ (green), respectively. Above 400 nm the fits have been scaled to the observed spectrum. The deviations of the fits from observed spectrum at $\lambda < 400$ nm reflect the influence of the 400 nm step size. For $I_{EBL} = 0$ the step size of the model fit is smaller while for $I_{EBL} = 8\ 10^{-9}$ erg cm$^{-2}$ s$^{-1}$ sr$^{-1}$ Å$^{-1}$ it is larger than the observed one. For an intermediate value of $I_{EBL} = 3\ 10^{-9}$ erg cm$^{-2}$ s$^{-1}$ sr$^{-1}$ Å$^{-1}$ a good agreement is reached. Fig. 7 also demonstrates the use of the Ca II H and K lines for the determination of the EBL. Again, for $I_{EBL} = 0$ the model predicts too shallow lines while for $I_{EBL} = 8\ 10^{-9}$ erg cm$^{-2}$ s$^{-1}$ sr$^{-1}$ Å$^{-1}$ the predicted lines are deeper than the observed ones; the

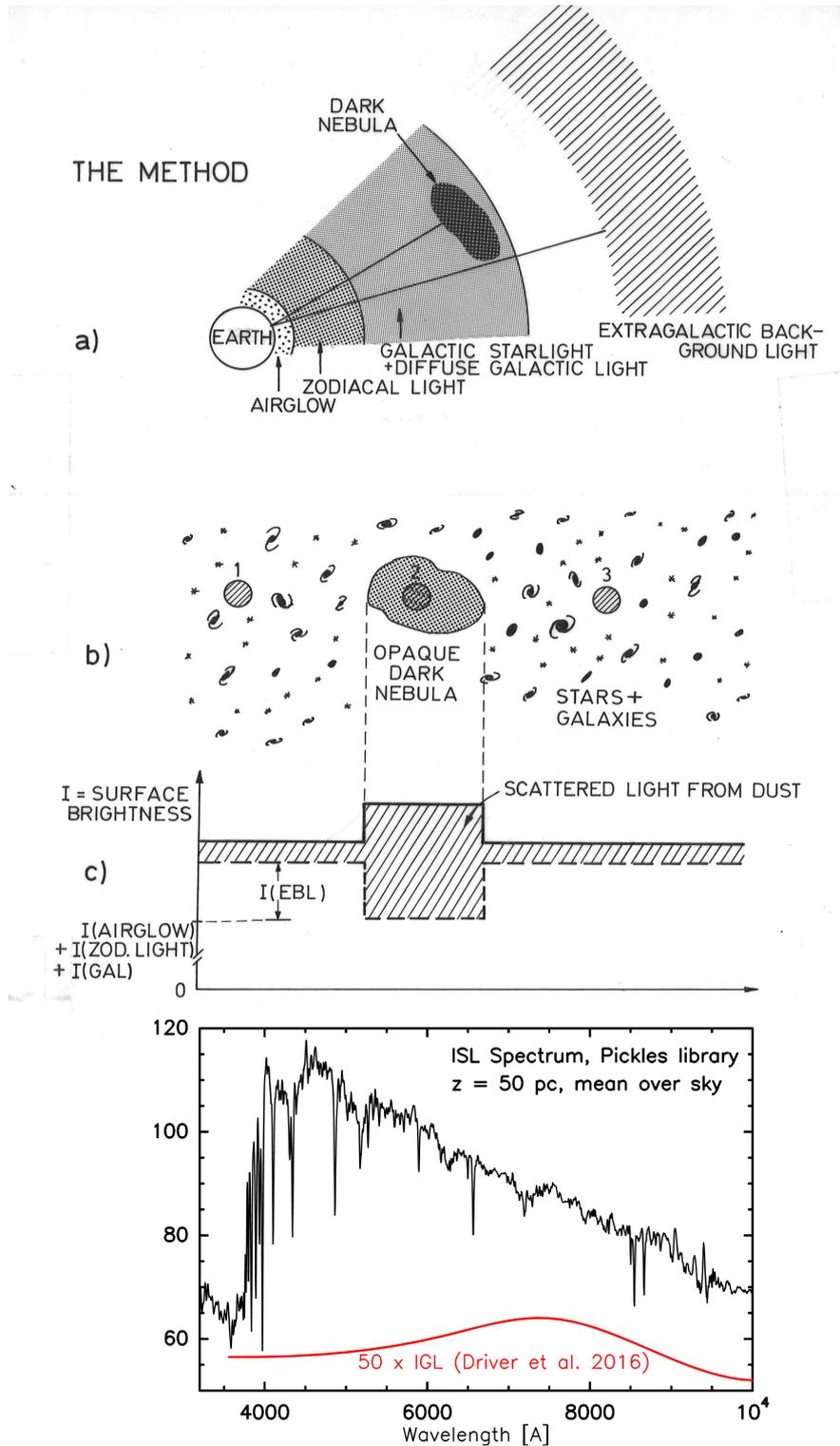

**Figure 5.** The principle of EBL measurement with the dark cloud shadow method, see text for details. (a) an overview of the geometry of the night sky components involved; (b) sketch of the view on the sky with three observing positions indicated. (c) If the scattered light from the interstellar dust were zero the surface brightness towards the cloud would be darker by the amount of the EBL intensity (dashed line). In reality, the scattered light is not zero, however, and the diffuse scattered light (shaded area) is particularly strong in the direction of the dark cloud making it brighter than the surrounding sky. (d) The spectral shape of the integrated Galactic starlight (with Fraunhofer lines) and the expected shape of the EBL (smooth spectrum, red line) are different. The synthetic spectrum of integrated Galactic starlight, mean over the sky and smoothed to a resolution of 1.1 nm, has been calculated for a virtual observer at a z-distance of 50 pc off the Galactic plane near the Sun; the unit is $10^{-9}$ erg cm$^{-2}$ s$^{-1}$ sr$^{-1}$ Å$^{-1}$ (see [62], Appendix A).

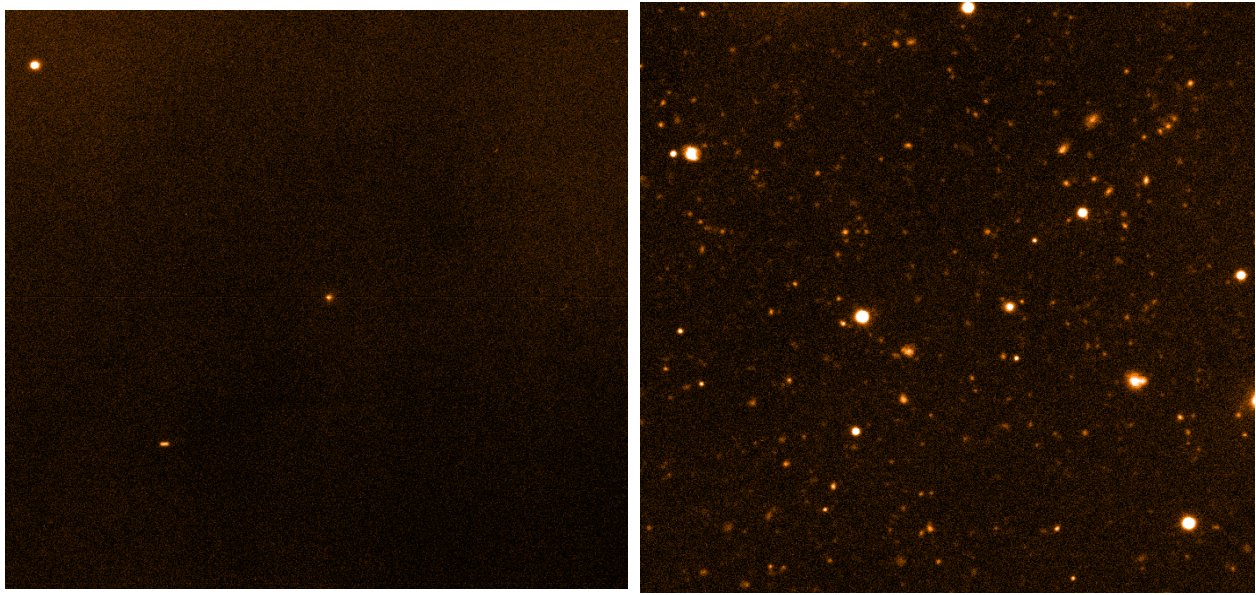

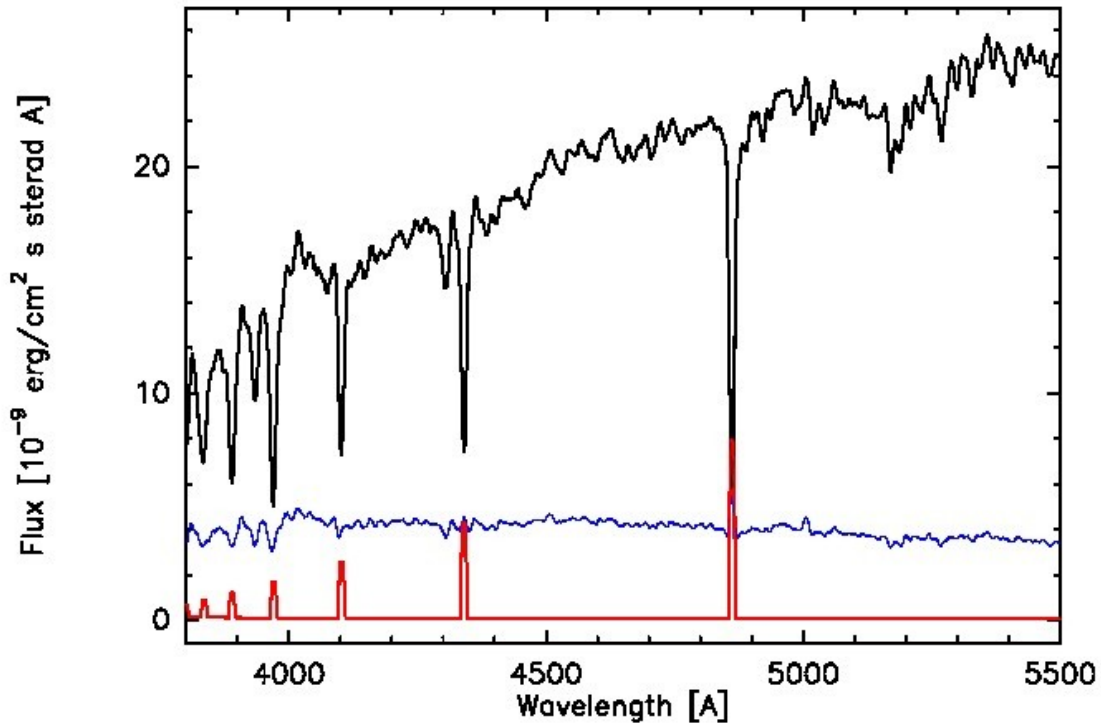

**Figure 6.** Examples of observed fields in dark nebula Lynds 1642 and its surroundings. *upper left:* opaque position in the centre of the cloud with visual extinction of >15 mag. *upper right:* example of a background sky position with good transparency, with visual extinction of ∼0.1 mag. The limiting magnitude of these ESO's Very Large Telescope B band images is ∼26.5 mag and they cover an area of ∼3' × 3' ; they have been scaled exactly the same way. Most of the objects seen in the right hand image are galaxies; the three objects in the left hand figure are stars in front of the cloud; *bottom:* The spectrum ΔI(λ) (on − off) of the opaque central position (on) minus transparent surroundings (off) is shown as black line. The blue line represents the scattered light spectrum and the red line the direct line-of-sight ionized gas emission-line spectrum at the off positions.

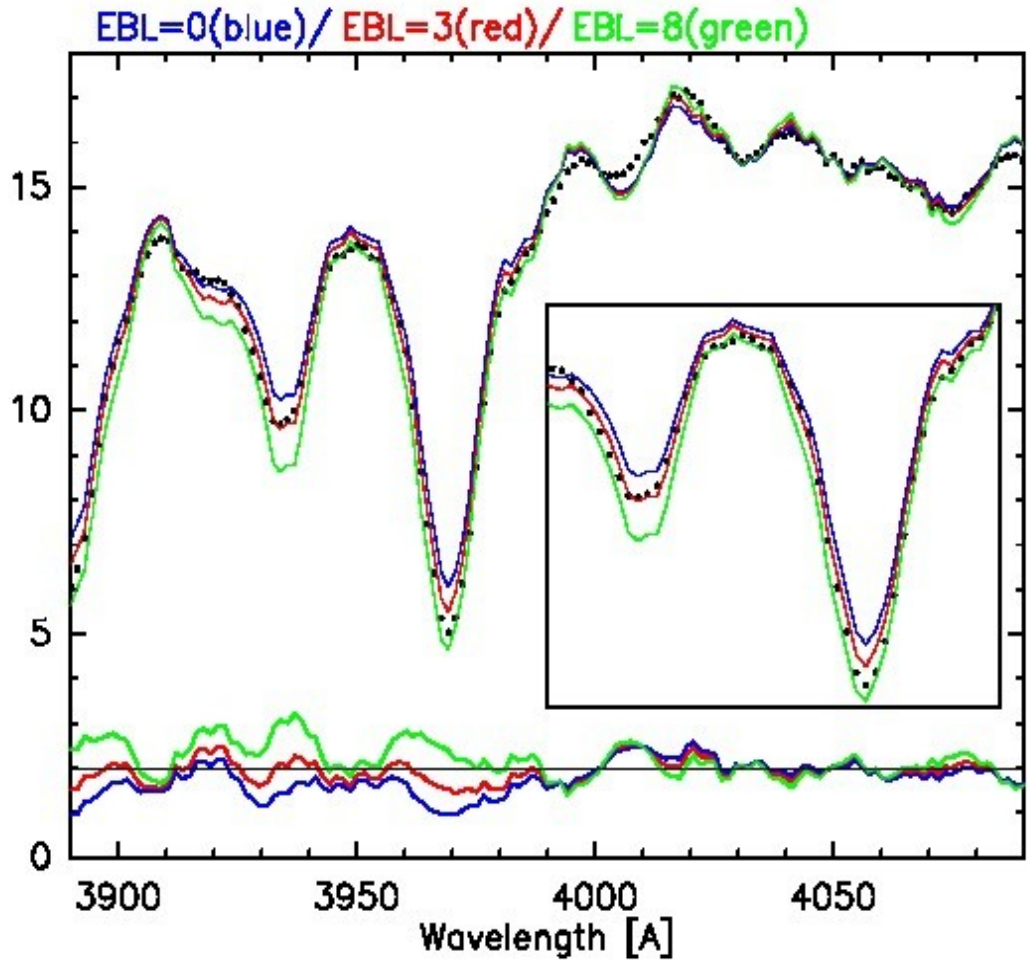

**Figure 7.** Demonstration of using the 400 nm discontinuity and the CaII H and K line depths for the determination of the EBL. The 389–409 nm on – off spectrum of the opaque position (black dots) is fitted for three assumed values of the EBL, $I_{EBL}$ = 0 (blue), 3 (red) and 8 (green) $10^{-9}$ erg cm$^{-2}$ s$^{-1}$ sr$^{-1}$ Å$^{-1}$; the fits are shown superimposed on the observed spectrum. At bottom the residuals are shown for each fit using the same colour codes and shifted by +2 units; the zero level is shown as black line. In the insert the range of the CaII H and K lines is shown at magnified scale. For the fitting procedure the idl (http://www.exelisvis.com/ProductsServices/IDL.aspx) programme MPFITFUN (www.physics.wisc.edu/˜graigm/idl/fitting.html) was used.

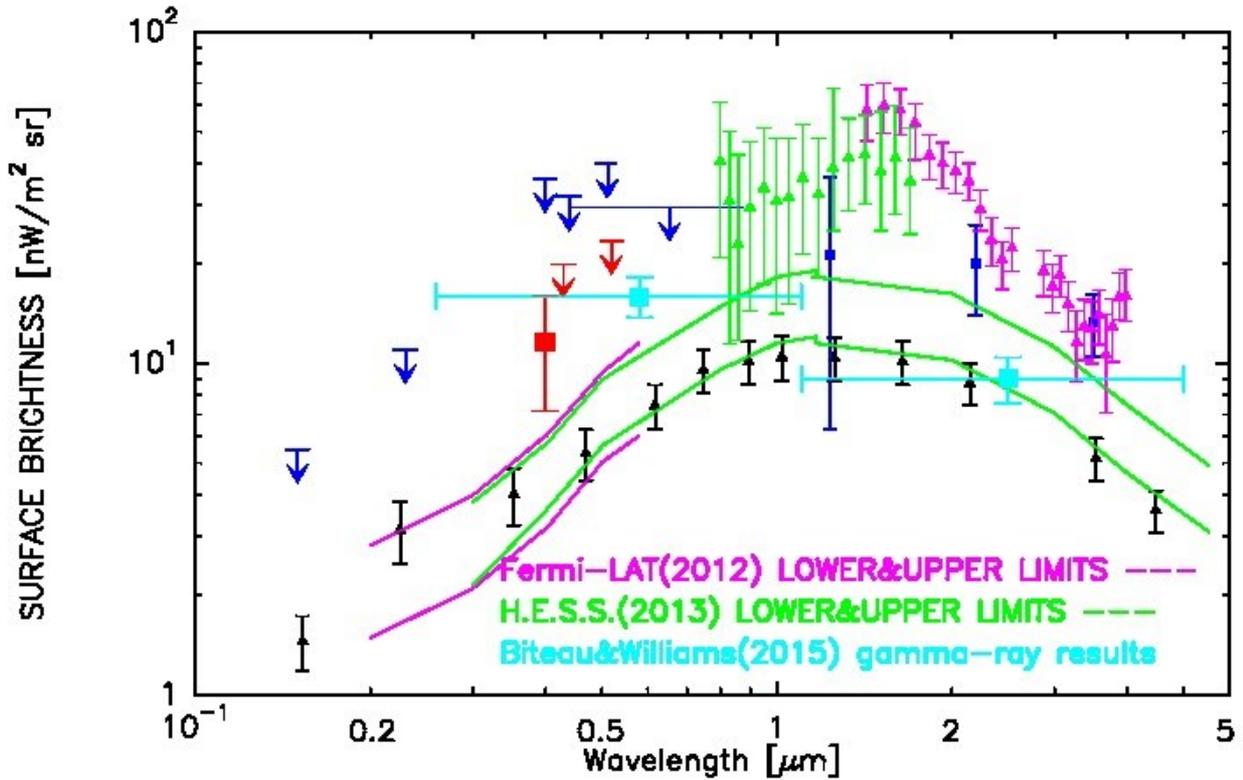

**Figure 8.** Selection of EBL measurements, upper and lower limits. *Integrated Galaxy Light (IGL):* 0.15–4.5 µm, Driver et al. [38], black triangles. *Photometric EBL measurements and upper limits:* red square with 1σ error bars at 400 nm and 2σ upper limits at 430 and 520 nm, Mattila et al. [62]; dark blue squares and arrows:150 and 230 nm, Hamden, Schiminovich, & Seibert [68], Akshaya et al. [69]; 400 nm, Mattila [31], 440 nm, Toller [43]; 512 nm, Dube et al. [51], the latter three according to re-discussion in Leinert et al. [36]; New Horizons upper limit between 440 - 860 nm, Zemcov et al. [46]; 1.25, 2.2 and 3.6 µm, Levenson, Wright, & Johnson [71]; near-IR spectrophotometry, 1.4 – 4.0 µm, magenta triangles, Matsumoto et al. [74] and 0.8 – 1.7 µm, green triangles, Matsuura et al. [76]. *Results from γ-ray attenuation:* upper and lower limits between 0.20–0.58 µm from Fermi Gamma-ray Space Telescope Large Area Telescope (Fermi-LAT) [106], magenta lines; between 0.30–5.5 µm from High Energy Stereoscopic System (H.E.S.S.) [105], green lines; results of Biteau & Williams [108] are shown as light blue crosses.

**Table 2.** Compilation of optical EBL observing methods and some recent results. $I_{tot}$ is the total sky, $I_{ZL}$ the zodial light and $I_{DGL}$ the diffuse galactic light intensity. HST stands for Hubble Space telescope, NH for the New Horizons space probe and AU for astronomical unit.

| Method and Reference | Wavelength [nm] | $I_{EBL}$ [nWm$^{-2}$sr$^{-1}$] | Comments |
|---|---|---|---|
| Galaxy counts [38] | 400 | $4.6 \pm 0.8$ | including all galaxies |
|  |  | $2.4 \pm 0.8$ | excluding galaxies $B < 22^m$ |
| Dark cloud | 400 | $11.6 \pm 4.4$ [a] | excluding galaxies $B < 22^m$ |
| shadow [62] | 400 | $13.6 \pm 4.5$ [a] | including all galaxies |
|  | 430 | $\leq 20.0$ [b] | excluding galaxies $B < 22^m$ |
|  | 520 | $\leq 23.4$ [b] | excluding galaxies $B < 22^m$ |
| $I_{tot}$ minus $I_{ZL}$ [56] | 300 | $\leq 42$ [c] | $I_{tot}$ from HST, $I_{ZL}$ from |
|  | 555 | $\leq 110$ [c] | ground-based observations, |
|  | 814 | $\leq 120$ [c] | $I_{DGL}$ from modelling |
| $I_{tot}$ beyond | 440-870 | $\leq 29.6$ [d] | $I_{tot}$ from NH at 7-17 AU, |
| zodiacal cloud [46] |  |  | $I_{DGL}$ from modelling |
| $\gamma$-ray absorption [108] | 260-1200 | $15.9 \pm 2.2$ | from blazar absorptions |

*a* 1 σ statistical error; scaling uncertainty is +20%/−16%

*b* 2σ upper limits;scaling uncertainty is +20%/−16%

*c* 2σ upper limits adopted as the final result of [52–56]; quoting from[56]: *"The EBL23 results we obtain are roughly 1-2σ detections and can be quoted as upper limits at the +2σ values on their own."*

*d* 2σ statistical error plus systematic error intermediate value, $I_{EBL}$= 3 10$^{-9}$ erg cm$^{-2}$ s$^{-1}$ sr$^{-1}$ Å$^{-1}$ , is seen to give the best fit. From least-square fitting the following main results were obtained in [62]:
(1) The EBL has been detected at 400 nm at 2.6-σ level, $I_{EBL}$ = 2.9 ± 1.1 10$^{-9}$ erg cm$^{-2}$ s$^{-1}$ sr$^{-1}$ Å$^{-1}$ or 11.6 ± 4.4 nW m$^{-2}$ sr$^{-1}$, scaling uncertainty +20%/-16%; (2) At 430 and 520 nm significant 2-σ upper limits were determined. This EBL value and the upper limits are shown in Table 2 and in Fig. 8 as the red solid square with 1-σ error bars at 400 nm and as 2-σ upper limits at 430 and 520 nm.

## 3. EBL in the UV and near-IR domains, and remarks on EBL fluctuations

Galaxies, intergalactic stars or any other light sources that give rise to the optical EBL will likely contribute to the background light in the adjacent UV and near-IR wavelength bands as well. We will briefly review some of the recent observational results in these wavelength bands and their relevance for the optical EBL.

In the near-UV (λ ≈ 230 nm) and far-UV (λ ≈ 150 nm) the zodiacal light foreground, that plagues the optical and near-IR EBL measurements is absent or much reduced. However, other difficulties appear [63–65] . The Galaxy Evolution Explorer (GALEX) [66] has provided a comprehensive survey of the near-UV and far-UV sky. Also, a rocket experiment [67] has covered a large fraction of the sky at 174 nm. The minimum sky brightness seen by GALEX is ∼ 300 − 400 (far-UV) and ∼ 600 photons cm$^{-2}$ s$^{-1}$ sr$^{-1}$Å$^{-1}$ (near-UV) toward the North and South Galactic Pole, respectively [68,69]. While these values still contain airglow as 'likely the dominant contributor' [68] they can be used to set upper limits to the EBL: $I_{EBL}(\lambda)$ <∼6−8 and $<\sim 12$ nW m$^{-2}$ sr$^{-1}$ in the far-UV and near-UV band, respectively. The scattered light from dust has been estimated to be $\sim 1.2$ - $1.8$ (far-UV) and $\sim 1$ nW m$^{-2}$ sr$^{-1}$ (near-UV) [69], thus lowering the likely EBL upper limits to $I_{EBL}(\lambda) <\sim 4 - 7$ and 11 nW m$^{-2}$ sr$^{-1}$ in the far-UV and near-UV, respectively (see Fig. 8). From the rocket experiment [67] data an EBL intensity of $4 \pm 2$ nW m$^{-2}$ sr$^{-1}$ at 174 nm was estimated.

The near-IR diffuse background sky brightness has been studied by several groups using the Cosmic Background Explorer's Diffuse Infrared Background Experiment (DIRBE) data in combination with The Two Micron All Sky Survey star catalogue (see [70–72] for reviews). Most of these results are consistent with the three selected values shown in Fig. 8 as blue squares at 1.25, 2.2 and 3.6 μ [71].

On the other hand, using data from the Infrared Telescope in Space spectrometer Matsumoto et al. [73, 74] have announced detection of emission at $\lambda = 1.4 - 4$ μm up to three times as large as the photometric values [71]. A similarly large excess has also been found using more recent space-borne spectrophotometric data at $1.8 - 5.3$ μm [75] and at $1 - 1.7$ μm [76]. As origin of this suspected large near-IR EBL the highly redshifted ($z >\sim 7 - 20$) UV radiation of the first generation of stars (so-called Population III) has been suggested; their strong UV radiation above the Lyman limit at $\lambda_{Ly} = 0.0912$ μm is redshifted to $\lambda > (1 + z)\ 0.0912$ μm or $\lambda >\sim 1$ μm while the radiation below this wavelength has been absorbed by intergalactic hydrogen along the path to the observer.

All these near-IR EBL determinations rely heavily on the model estimates of the zodiacal light intensity [77] or [78,79] that, besides starlight and diffuse galactic light, has to be subtracted from the observed total diffuse sky brightness. It has been pointed out [70,80] that a possible alternative explanation for the suspected large near-IR background excess could be an insufficient zodiacal light correction resulting from inadequacy of the models. Besides the zodiacal light also the diffuse galactic light modelling introduces substantial uncertainty.

Because of the inhomogeneous distribution of galaxies and other luminous matter in the Universe the diffuse sky backgrounds show fluctuations. Observationally, the fluctuations have an advantage over the mean EBL measurements, namely that any spatially uniform foreground components, such as the zodiacal light, are cancelled out. However, the rms amplitude of the fluctuations is expected to be only a minor fraction, from $\sim 5$ up to $\sim 20$ per cent, of the mean EBL intensity. Therefore, the spatially varying foreground components, such as the instrumental aureoles and ghosts caused by stars and bright galaxies as well as the fluctuations of the diffuse galactic light, do strongly influence the EBL fluctuation studies.

In the optical domain, Shectman [81, 82] pioneered the studies of fluctuations as a potential source of cosmological information and interpreted the observed angular power spectrum in terms of clustering of unresolved galaxies at modest redshifts, $z = 0.2 - 0.6$. In the UV at $\lambda = 1350 - 1900$ Å the angular power spectrum of sky fluctuations was measured and interpreted again in terms of galaxy clustering [83]. In a subsequent paper the same team [84] argued, however, that their search and the earlier optical search in [81,82] had detected only the fluctuations of the reflected Galactic starlight from dust, well-known by that time as 'IRAS infrared cirrus' [85].

In the near-IR domain the study of the background sky fluctuations has developed into a wide research field on its own right, see [34]. The motivation for these efforts has largely been the prospect of discovering the radiation of the first stars and galaxies [86–94]. Other light sources have been discussed as potential contributors to fluctuations: (1) the intra-halo light, that is the light of stars in galaxy halos and intergalactic space inside clusters and groups of galaxies [94,95]; (2) faint, dwarf galaxies at intermediate redshifts, e.g. [96,97]; and (3) at large angular scales, $\theta > 4$ arcmin, the foreground diffuse galactic light may contribute substantially, depending, however, on the greatly varying amount of dust along the line of sight [98–100]. The different competing interpretations are still hotly debated.

In a representative recent study [99], a large area of $\sim 15 \times 15$ arcmin$^2$ has been observed in the Hubble Space Telescope legacy program in five wavebands, $\lambda = 0.606-$ 1.6 µm, and analyzed with respect to the background fluctuations. Beyond the shot noise domain at $\theta < 10$ arcsec the angular power spectra showed an excess power. It was suggested to arise, at scales up to $\theta \sim 200$ arcsec, from the intra-halo light associated with faint low-redshift ($z \sim 0.5-1$) dwarf galaxies inside clusters and groups of galaxies. When interpreted in terms of the galaxy model [96] the contributions of intra-halo light to the mean EBL intensities ranged from 0.13 +0.08/-0.05 nW m$^{-2}$ sr$^{-1}$ at 0.60 µm to 0.54 +0.58/-0.31 nW m$^{-2}$ sr$^{-1}$ 1.6 µm. These values are only 2 − 5 percent of the integrated light of galaxies at these wavelengths and $\sim 1/10$th of the intra-halo light intensity found in a previous study using more limited rocket-borne fluctuation data at 1.1 and 1.6 µm [12]. At still larger scales, $\theta > 200$ arcsec, the foreground diffuse galactic light fluctuations were found to be the dominating component.

## 4. A different approach: EBL measurement using absorption of γ-ray radiation from background sources

For γ-ray radiation of high (HE, 0.1 - 100 GeV) and very high energy (VHE, > 100 GeV) the Universe ceases to be transparent. While passing from a γ-ray source (e.g. a blazar) to the observer on Earth the γ-ray photons interact with the photons of the EBL field. This process leads to creation of e+ , e− pairs and lower-energy photons [101–103]; see the cartoon of Mazin and Raue, Fig. 9. Because the harder γ-rays are attenuated more strongly and are even cut out above a certain energy limit, this absorption process changes the observed γ-ray spectrum of blazars in a characteristic way (see Fig. 9) and provides, thereby, a useful indirect method to probe the EBL. This way, at first the mid-IR EBL [104] and more recently also the optical EBL [105–108] have been probed; for a review see e.g. [109–111].

The peak cross section of the pair production process occurs when the energies of the γ-ray and optical-IR photon fulfil the condition: $E_\gamma \times E_{opt-ir} \approx 4 \, (m_e c^2)^2 \approx$ 1 MeV$^2$ depending, however, strongly on the angle between the photon propagation directions. Thus, to probe the EBL at $E_{opt-ir} \approx 1$ eV, corresponding to $\lambda \sim 1.24$ µm, γ-rays with $E_\gamma \approx 1$ TeV will be most effective; to probe the optical EBL one requires sensitive γ–ray observations at $E_\gamma << 500$ GeV.

The energy dependence of the cross section is, however, very wide. This leads to a limited spectral resolution so that even large variations in the EBL spectrum are smoothed out over a wide range of γ-ray energies or, in other words, the EBL intensity at a given wavelength has an effect on the γ-ray opacity over a wide range of $E_\gamma$ . Also, while the γ-ray absorption method naturally is free of the problems caused by foreground sky components, it does include uncertainties of the intrinsic spectral energy distributions of the blazars used as probes for the γ-ray absorption effects.

The intergalactic radiation density has been derived from γ-ray attenuation measurements of blazars by the ground-based Cerenkov telescopes, High Energy Stereoscopic System (H.E.S.S.), Major Atmospheric Gamma Imaging Cherenkov Telescopes (MAGIC) and Very Energetic Radiation Imaging Telescope Array System (VERITAS), sensitive in the 0.2–20 TeV (VHE) range, as well as by the Large Area Telescope (Fermi-LAT) aboard the Fermi Gamma-ray Observatory, sensitive in the 1–500 GeV (HE) range. Results based on extensive data sets from Fermi-LAT [106] and H.E.S.S. [105,112] are shown in Fig. 8 as upper and lower boundaries to the allowed range of EBL values. Similar results have been published by the MAGIC [113,114] and VERITAS collaborations [115]. Fermi-LAT probes with high sensitivity the UV–to–optical ($\lambda \approx 200$–600 nm) EBL while the ground-based VHE telescopes cover the wider range from optical to far-IR, $\lambda \approx 0.3$–100 µm.

The EBL spectrum as presented by the H.E.S.S. and Fermi-LAT collaborations is the result of fitting a template spectrum to the $\gamma$-ray attenuation data. The template comes from a model that is obtained by summing up empirically–constrained galaxy populations spectra [107,116,117]. With a normalization factor as the only free parameter for the $\gamma$-ray derived EBL spectrum its shape is forced to closely follow the spectrum of the integrated light of galaxies. Because of the broad wavelength response function of $\gamma$-ray absorption vs. optical and infrared EBL photons the intensity maximum at $\sim 1$ µm dominates the normalization factor over the whole UV–near-IR range 0.2–5 µm and it is difficult to detect the possible deviations from the spectral shape given by the integrated light of galaxies using this 'model template approach'.

Recently, Biteau & Williams [108] (see also H. E. S. S. Collaboration et al. [112] and Moralejo et al. [118] for the MAGIC collaboration) have taken a more general approach. They abstain from assuming *a priory* a template spectrum shape defined by the integrated light of galaxies. Using $\gamma$-ray observations only they were able to derive a 'free-standing' EBL spectrum which covers the wavelength range 0.2–100 µm with four spectral bins, each with a width of $\sim$ half a decade. Two of these spectral bins, 0.26–1.2 µm and 1.2–5.2 µm, are displayed in Fig. 8.

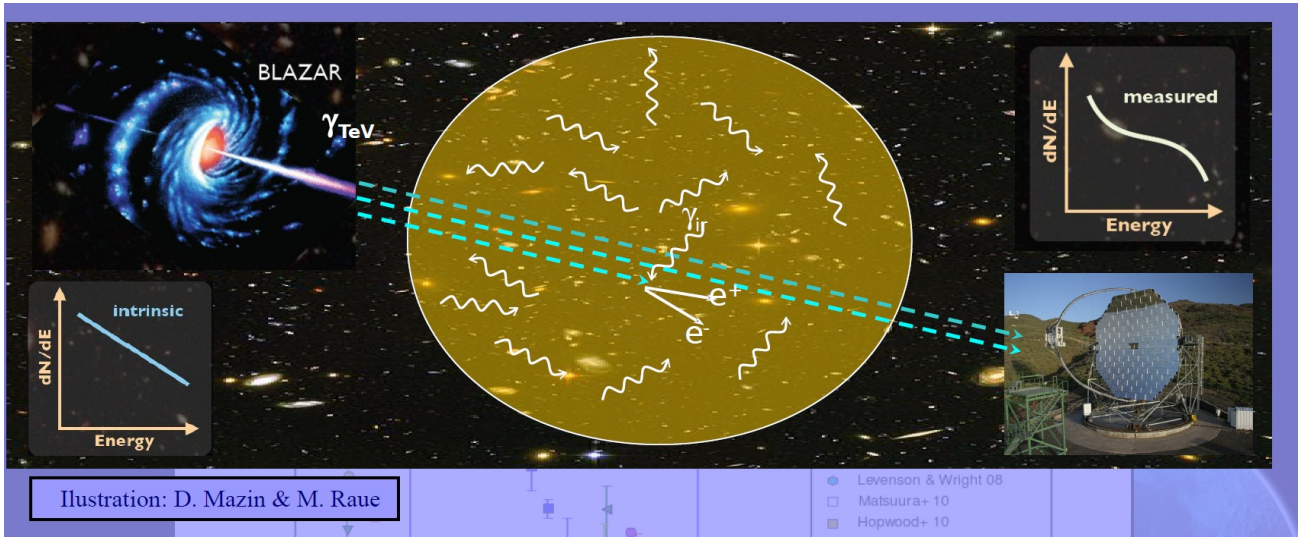

**Figure 9.** Demonstration of the attenuation of $\gamma$-ray radiation from a blazar by intergalactic optical-to-IR radiation field. (copyright D. Mazin & M. Raue)

## 5. Integrated light of galaxies: counting galaxies

The obvious contribution to EBL is the summed total light of galaxies in the universe, and this integrated light is sometimes confused with EBL itself. While the integrated light of galaxies may or may not be close or equal to EBL, it does serve as a convenient lower limit. We briefly review the galaxy number counts and the level of their integrated light, and compare it to EBL measurements as discussed in Sections 2 – 4.

In general, the total light of galaxies, at a given frequency $v_0$ and zero redshift, can be expressed as

$$IGL = \frac{1}{4\pi} \int_0^{z_f} \frac{dl}{dz} \frac{F(\nu, z, type)}{(1+z)^3} \, dz,$$

where $F(v, z, type)$ is the galaxy emissivity at redshift z and frequency $v = v_0 (1 + z)/(1+z_0)$, and where the integral must be summed over different types of galaxies. The F term contains the details of spectral energy distributions for given galaxy type, as well as type dependent luminosity functions (number of galaxies per luminosity bin in a given volume). The evolution of the galaxy population, or the star formation history, is introduced via redshift dependent luminosity functions, as well as the formation epoch of the galaxies $z_f$. Cosmological model dependence enters due to varying volumes in different geometries through the term $dl/dz$ where $l$ is the distance.

Observationally, the integrated light of galaxies can be summed up from the number counts of galaxies,

$$IGL = \int N(m) \, 10^{-0.4m} \, dm$$

where $N(m)$ is the number of galaxies observed per magnitude (brightness) bin. Since N (m) contains the same $F(v, z, type)$ as above, galaxy counts have been used for decades as a cosmological test and as a constraint on galaxy evolution scenarios [119–122]. This observational approach provides the integrated light down to a given brightness, though to arrive to the total integrated light value an extrapolation of the counts is needed.

In the context of EBL it is crucial to note that the integrated light value will converge mathematically if the slope α of the counts, expressed in the form $\log_{10} N(m) = \alpha m + const$, is α < 0.4. It turns out that the galaxies with brightnesses close to these limiting cases produce the bulk of the observed integrated ligh. This is demonstrated in Fig. 10 where deep galaxy counts are shown (left panel) together with their corresponding contribution to the integrated light of galaxies (right panel). It is also clear that extrapolations to fainter galaxies beyond current detection limits will not contribute appreciably to the integrated light. In other words, it appears that the integrated light is essentially resolved. The most recent integrated light values from UV to near-IR (0.1 - 5 µm), from a compilation of the deepest available galaxy counts [38] are shown in Fig. 8 with black triangles.

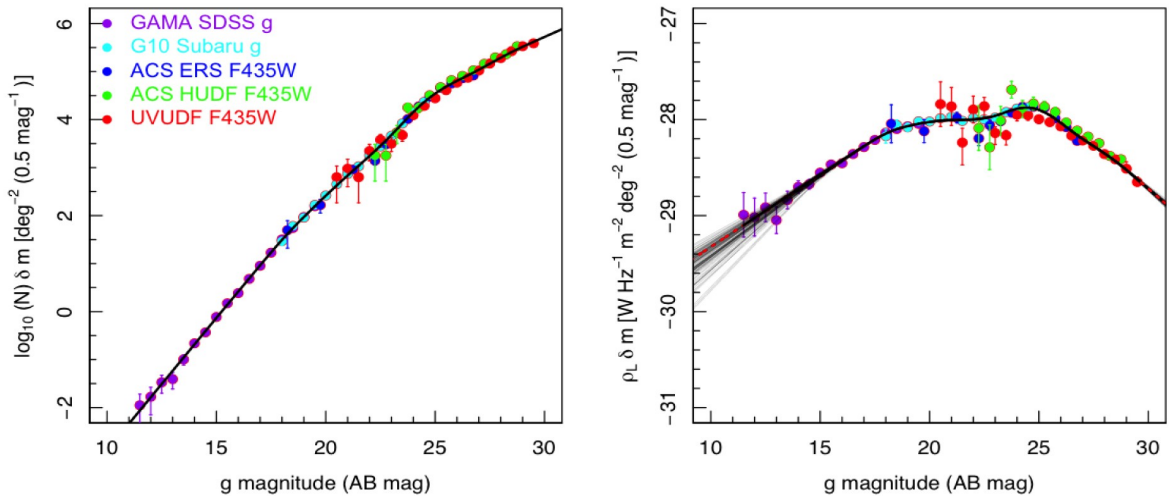

**Figure 10.** The left panel shows a compilation of deep galaxy number counts in the blue optical g-band. The contribution of each magnitude (brightness) bin to the overall luminosity density from these galaxies is plotted on the right. The contribution to the integrated light from galaxies fainter than g ∼ 25 mag, where the count slope flattens, becomes progressively smaller. (Adopted from Driver et al. [38], Fig. 2, courtesy of Astrophysical Journal).

## 5.1 Does the integrated light of galaxies account for the total EBL?

In Fig. 8 and Table 2 the integrated light from galaxy counts is compared with a selection of EBL measurements and upper and lower limits (colour symbols and lines). It is worth noting that in this comparison any bright end limits need also to be taken into account. For example, the recent dark cloud method [62] excludes galaxies brighter than ∼ 22 mag, which according to the galaxy counts [38] would contribute ∼2.2 nW m$^{-2}$ sr$^{-1}$ of the total light produced at 400 nm, resulting in IIGL (m ≥ 22 mag) = 2.4 ± 0.8 nW m$^{-2}$ sr$^{-1}$ which differs from the EBL value [62] IEBL (m ≥ 22 mag) = 11.6 ± 4.4 nW m$^{-2}$ sr$^{-1}$ by a significant factor (see Table 2). One possible concern in this comparison is cosmic variance, field to field variations due to galaxy clusters and structures within the galaxy counts used for integrated light estimates or, likewise, diffuse cluster light in EBL measurements. However, the statistical effect is shown to be at most at a 20% level even in smallest used galaxy count field areas [38,123], where the effect would be the largest. On the other hand obvious clusters are avoided when selecting EBL fields, and the effect of any nearby ones can be evaluated [62].

In UV the EBL estimates (upper limits) are by a factor of ∼ 4 higher than the corresponding values of the integrated light of galaxies as displayed in Fig. 8.

The photometric near-IR results suggest that the near-IR EBL may exceed the integrated light of galaxies by a moderate amount, however not more than a factor of ∼ 2 − 3 [71]. On the other hand, Matsumoto et al. [73, 74] have claimed an excess emission at λ = 1.4 − 4 µm up to six times as large as the integrated light of galaxies. Similarly large excess has also been found independently at 1.8 − 5.3 µm [75] and at 1 − 1.7 µm [76]. As shown in Fig. 8 these excess values are, within the error bars, still marginally compatible with the photometric values [71].

In conclusion and as demonstrated by the data in Fig. 8, the photometric measurements in the UV (150 − 250 nm) and near-IR (1 − 5 µm) suggest that the mean EBL intensity exceeds the integrated light of galaxies, by a factor of two or even more.

There remain, however, substantial uncertainties in these EBL measurements because of the difficulties with foreground subtraction.

The lower limits to EBL from the γ-ray attenuation results, ascribed to H.E.S.S. and Fermi-LAT in Fig. 8, are seen to agree with the integrated light from galaxy counts while their upper bounds do allow a modestly higher (by up to 70%) EBL intensity as well. The good agreement of spectral shapes of the integrated light of galaxies and EBL is, however, a direct result of the underlying assumption of an integrated-galaxy-light template spectrum. For the EBL values of Biteau & Williams [108] which are independent of spectral template assumptions there is a good agreement with the integrated light of galaxies for the three bins at 1.2–5.2 μm, 5.2–23 μm, and 23–103 μm, the bin at 0.26–1.2 μm showed an excess of 4.7±2.2 nW m$^{-2}$ sr$^{-1}$ over the appropriately weighted integrated light mean value. This excess over the integrated light value, as well as over the Fermi–LAT and H.E.S.S. EBL ranges, is also well demonstrated by Fig. 8. We note that while the γ-ray based EBL value [108] of 15.9±2.2 nW m$^{-2}$ sr$^{-1}$ for the bin 0.26–1.2 μm corresponds to the total EBL, the photometric EBL measurement [62] at 400 nm does not include the contribution of the bright galaxies with g ≤22 mag. Including that contribution, the photometric EBL value increases from 11.6 ± 4.4 nW m$^{-2}$ sr$^{-1}$, as shown in Fig. 8, to 13.6 ± 4.5 nW m$^{-2}$ sr$^{-1}$ (see Table 2) which is in good agreement with the γ–ray value [108].

## 6. Possible sources of EBL beyond galaxy counts

It has been seen in the previous section 5.1 that most determinations of the EBL are in excess of the summed and extrapolated galaxy light by up to a factor of ∼ 2 − 3 in the blue optical and the near-IR light. While it may be difficult to rule out that the large foregrounds of EBL are still not subtracted correctly, the variety of independent methods in measuring the optical and near-IR EBL do motivate for seriously exploring causes of the discrepancy or potential physical processes responsible for the difference. Possible extra light sources that are known to exist include light from Milky Way halo, from the outskirts of galaxies, or from intergalactic stars in galaxy clusters and groups. The contribution of such light sources may in certain objects or environments be substantial.

### 6.1 Unresolved starlight and the Milky Way halo

Sky background light as seen by an observer inside the Galaxy contains contributions from unresolved stars, diffuse emission from gas, and scattered light from dust. The light from unresolved stars can be estimated by using the deep star counts as presented e.g. in [124]. While the number of stars at B ≈ 20 mag at middle and high galactic latitudes is roughly equal to galaxies, the stellar number count slope is much shallower than that of galaxies beyond 20 mag. Thus the contribution of stars drops to less than 1% of integrated light of galaxies soon after the inflection point of the galaxy counts (see Fig. 10) over the ≈ 25 − 28 magnitude range.

Extended ultraviolet (λ ≈ 150 − 250 nm) halos have been observed around several edge-on late-type galaxies out to 5–10 kpc from the mid-plane, e.g. NGC 891, NGC 5907 [125–127]. They have been ascribed to thick dust disks that scatter light from stellar disks of the galaxies. The existence of substantial amounts of dust in the Milky Way halo at, say |z| > 2 kpc, is not known. Note also that in the case of the dark-cloud method it would be eliminated from the EBL measurement by virtue of the spectral separation since any such dust would be illuminated by the Milky Way radi-ation field that has closely the same absorption line spectrum as the scattered light of the cloud. Gas emission from Milky Way halo, on the other hand, is expected to be almost entirely in the form of line emission, mostly the hydrogen Balmer lines [128,129]. Since the continuum is very weak, contributions to broad or even narrow band photometric measurements are negligible. However the emission lines, if included in the wavelength slots used for EBL mesurement, have to be taken into account. Avoiding slots with the strong lines Hα, Hβ, Hγ, their influence was found to be small for the result in [62].

## 6.2 Missed light at the outskirts of galaxies

Measuring the faint outer structures of galaxies is very challenging. An accurate determination of a blank "sky" region around the target, or somewhere on the CCD detector, is required to arrive at an absolute result. Recent work to gain insight into galaxies' evolutionary past by studying their stellar halos at extremely faint levels of $>\sim 31$ mag/sq.arcsec [130–132], se Fig. 11 left panel, illustrate both the instrumental difficulties (scattered light in the optics, flat fielding, etc.) as well as ambiguities in the photometric techniques. Nevertheless, despite detections of spectacular wide and extended structures of faint diffuse light around nearby galaxies in these studies, the actual fraction of total "missing" light is only of the order of few percent.

However, it may be that when measuring the light of distant galaxies the missing fraction of light is larger due to the photometric techniques employed. It is clear that not all light of a faint distant galaxy is captured by fixed apertures or even adaptive 'total' magnitude aperture photometry. Totani et al. [133] studied the influence of light loss from the outskirts of galaxies, i.e. galaxy wings, via photometric modelling. They found that their 'best guess' integrated light from all galaxies in the Universe was up to 80% higher than the value from a simple integration of observed galaxy counts. In a detailed analysis of the Hubble Deep Field and another deep field Bernstein et al. [53, 54] concluded that a minimum of 20% of the flux of the faintest galaxies was contributed by galaxy wings at r > 1.4 $r_{iso}$, i.e. at radial distance often missed by typical photometric techniques. Furthermore, using a method called 'ensemble photometry' they estimated that the true flux from V > 23 AB mag galaxies in the Hubble Deep Field can be almost twice as much as that recovered by standard photometric methods. Benitez et al. [134] analysed the faint galaxy population in two Early Release Science fields as observed with the Hubble Space Telescope's Advance Camera for Surveys confirming these claims by finding an up to 50% loss of light of the faintest galaxies.

The galaxy photometry used by Driver et al. [38] for their integrated light values in the optical bands is, at the faint end, based on the Hubble Ultra Deep Field and other observations which are deeper than the data used by [53,54]. They are therefore less vulnerable to effects of galaxy wings, and were also designed to take better care of the galaxy-wing contribution. They conclude that galaxy wings should not have an effect of more than 20% on their faint galaxy fluxes, and that the effect even rather likely to be at < 10% level (Driver 2016, private communication). Independent tests were also run [62] on the Hubble Ultra Deep Field frames varying the sizes of the adaptive apertures at the faint limit, and found that while many galaxies *close to the detection limit* become significantly brighter when using larger apertures the overall effect is more at the 20% level and, moreover, this missing light contributed only 6% to the total integrated light over the full brightness range of galaxies in the field.

Thus, while it well may be that significant amounts of light are missed in typical galaxy photometric techniques, it might be difficult to imagine that those surveys where specific care is taken against such effects could still miss more than, say, 20% of the total light because of the galaxy outskirts. Moreover, an obvious counter-argument for large fractions of missing light at the outskirts of galaxies is the absence of large supernova populations in these areas. From the results of Bartunov, Tsvetkov, & Pavlyuk [136] one finds that only ~5–7% of the total number of supernova events occur at radii $r = (1.4 - 4)\, r_{25}$ .

## 6.3. Low surface brightness galaxies

Galaxies with very low surface brightness could escape detection altogether in galaxy counts, or their total brightness be significantly underestimated even if detected, due to various selection effects [137]. Hence it is in principle possible that these galaxies could contribute significantly to EBL while not adding to the integrated galaxy counts [11]). Low surface brightness galaxies exist with comparable number densities to normal galaxies [138] and with a wide variety of characteristics, from dwarfs to giants (e.g. Malin 1, the most famous extreme case [139]), and comprising both star-forming and red quiescent galaxies [140]. There has been a recent upsurge of interest in low surface brightness galaxies and so–called ultra diffuse galaxies due to the detection of surprisingly large populations of them in rich clusters of galaxies [141–143]. However, the total contribution of all the various types and classes of low surface brightness galaxies and ultra diffuse galaxies to the total integrated light of galaxies remains underwhelming. Notably, both deep optical surveys [145] and H I 21-cm surveys [146,147] over the past two decades have failed to detect populations of field low surface brightness galaxies significant enough to contribute in any significant way to the total luminosity density of the universe. Thus, the consensus appears to be that less than 20% of additional light to the integrated light could be contributed by the low surface brightness galaxies [144,145]. Nevertheless, due to the faintness of the population and the associated technical difficulties of detections, the exact characteristics of this galaxy population remains relatively poorly constrained, and e.g. Disney et al. [148] have recently argued that dim and/or dark galaxies might still be evading surveys if such galaxies were strongly clustered.

## 6.4 Diffuse light within galaxy clusters

Intra–cluster and intra–group light is a well established component among cosmic light sources. It originates from stars stripped off from galaxies in the cluster formation phase or in later interactions between galaxies or, perhaps, also from stars formed *in situ* in the intra-cluster gas [149].

Diffuse light between the galaxies was first noted by Zwicky [150] in deep photographs of the Coma Cluster. Wide-field CCD imaging has since then enabled deep surface brightness surveys of intergalactic light in many other clusters and groups e.g. [151–153]. The fraction of intra-cluster light of the total cluster luminosity varies between ~5–40% in nearby big clusters and those measured at intermediate redshifts, with the latter typically resulting in somewhat lower values [154–156]. Six clusters studied at redshifts of z ~ 0.8 − 1.2 resulted in intra–cluster light fractions of mere 1 to 5% [157]. In galaxy groups the fraction of diffuse light varies even more strongly from undetectable in loose groups to ~30–40% in many compact groups (see Fig. 11, right panel, and e.g. [158,159]).

Although the diffuse light fraction in individual clusters may be high, up to 40%, it is important to remember that rich clusters contribute only a few percent of the total cosmic starlight while ∼80% of the light comes from individual field galaxies or loose groups. Therefore, hardly more than 10% is added to the integrated light of galaxies by the contributions from intra-cluster and intra-group starlight. We also note that the intra-cluster supernova events account for <∼20% of the supernova rate for the clusters [160,161], in agreement with the estimates of the diffuse light fraction.

## 6.5 Other sources of diffuse light

Because of Lyman line and continuum absorption, the redshift range of light sources contributing to sky brightness at optical wavelengths is limited to redshifts z <∼ 8 thus excluding exotic primordial sources. In the blue optical regime at λ <∼ 450 nm any population III stars could not contribute, either, because the limit is $z$ <∼3.5.

Decaying or annihilating dark matter candidate particles, such as neutrinos, Weakly Interacting Massive Particles (WIMPs) and axions, have been proposed as possible sources of diffuse background radiation fields. On the other side, the EBL might qualify as an important discovery channel for the elusive dark matter particles; for a review see Overduin & Wesson [20, 162] and the update in [163]. Recently, the possibility has been discussed that the near-IR background *fluctuations* could partly originate from decaying axions with mass around 4 eV, located mainly in the halos of clusters and groups [164]. None of the three particle species have been found to produce enough radiation to qualify as serious contributor to the mean intensity of the EBL in UV [163], optical or near-IR [164] domain. Furthermore, axions and WIMPS should be strongly concentrated to the dark matter halos of clusters, groups and individual galaxies. Even if some part of the diffuse intra-cluster, intra-group or galaxy halo diffuse light were caused by decaying particles, instead of stars, this would not change the amount of diffuse light as determined by the observations. Only a smoothly distributed diffuse light component, present also in the general field outside the clusters, could have escaped the observations.

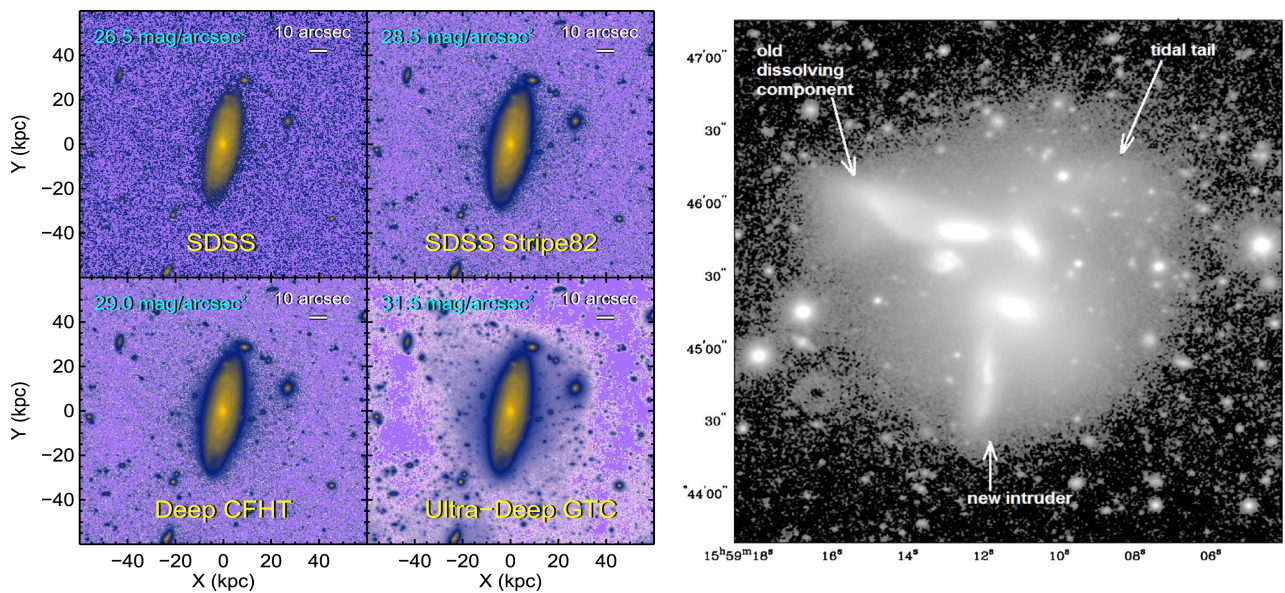

Figure 11. Left: The Sab type galaxy UGC00180 imaged to different depths, reaching from 26.5 mag/sq.arcsec in the upper left to the extremely faint surface brightness of ∼31.5 mag/sq.arcsec in the lower right hand panel; at ∼31.5 mag/sq.arcsec extended faint structures around the main body of the galaxy are revealed. Right: Seyfert's Sextet is an example of a compact group of galaxies with an extensive common diffuse halo. (The images are adopted from Trujillo & Fliri [130], Fig. 6 and Durbala et al. [135], Fig. 6, respectively, courtesy of Astrophysical Journal and Astronomical Journal.)

In summary, several of the sources discussed above, if carefully measured, could add light on top of the conventional integrated light by several tens of percentage points. However, none of these sources appear to be capable of explaining a substantial increase of the overall mean EBL beyond the integrated light as derived from galaxy counts.

## 7. Future prospects of EBL measurements

*Comparison of the photometric and γ-ray methods.* As has been described above there are two fundamentally different approaches to the measurement of the EBL intensity or the intergalactic radiation density at the optical, ultraviolet and infrared wavelengths:
(1) one can measure (spectro)photometrically the surface brightness of the diffuse background sky; the foreground components from the Earth's atmosphere, Solar system and Milky Way galaxy have to be eliminated or avoided as far as possible via the observing strategy; (2) the second method consists of measuring the attenuation that high-energy gamma-rays experience when passing through the intergalactic UV-to-IR radiation field; this surprising phenomenon of 'absorption of light by light' is based on the creation of e+ , e− pairs in the photon-photon interaction.

The two methods are in many respects complementary and connected with very different problems or advantages. The spectral energy distribution of the EBL can, *in principle*, be studied with good resolution by the (spectro)photometric method while the gamma-ray absorption method allows only estimates over broad spectral windows. The gamma-ray method, on the other hand, enables probing the EBL as function of distance up to substantial redshifts [116,117] which is not possible with the photometric method. Observationally, the photometric method is currently perplexed by difficulties of foreground elimination while the determination of the intrinsic spectral shapes of blazars remains a problem for the gamma-ray method [111,165,166].

*Future prospects of (spectro)photometric EBL measurements.* The rocket-borne Cosmic Infrared background Explorer (CIBER) high resolution ($\lambda/\Delta\lambda = 1120$) spectrometer [167] is expected to deliver a measurement of the depth of the Solar 854.2 nm Ca II Fraunhofer line in the zodiacal light spectrum. Using the approach of Dube et al. [50, 51] this will give an accurate measure of the absolute zodiacal light intensity and, when subtracted from the simultaneously measured total sky intensity, the combined value of EBL + diffuse galactic light. A second version of this series of rocket experiments [168,169] is under preparation and will have improved sensitivity and wavelength and area coverage.

*Platforms to the outer Solar system:* Sky background measurements with photometers aboard the Pioneer 10/11 and New Horizons planetary missions have demonstrated that the zodiacal light as seen towards the ecliptic poles and anti-solar directions drops by a factor of up to 100 from $r = 1$ to r > 3 astronomical units, thus offering an opportunity to measure the EBL (almost) free of zodiacal light contamination (see Section 2.3). It has been suggested [170] that the New Horizons extended mission, now already well beyond Pluto's orbit, should be used for an extensive optical broad band (as in [46]) and optical-UV spectral measurement of the EBL. Other ideas and suggestions for EBL measurements utilizing missions reaching outside the interplanetary dust cloud, in radial or polar direction, have been presented [29].

*An isotropic zodiacal light component?* Such a component, if present, was not contained in the zodiacal light models [77–79] which were based on the annual variability of zodiacal light intensity at the ecliptic poles. For discussion concerning a possible isotropic zodiacal light or Kuiper Belt Objects' contribution, see [15,171–173]. An isotropic zodiacal light component, even if small, would continue to bias the EBL measurements where EBL is derived as the difference between the total sky brightness and the zodiacal light. It has been argued that the minimum far-IR sky background sets a strict upper limit to the near-IR EBL which at 1 µm is much below the EBL [174].

*Diffuse galactic light remains an obstacle for photometric EBL measurements.* Diffuse galactic light is not a specific nuisance of the dark cloud method. So far, the diffuse galactic light estimates have often been derived using dust distribution and grain scattering models [46,74,76] or by correlating the optical and far-IR surface brightnesses and extrapolating to an assumed far-IR zero point [44]. However, for a secure measurement of the EBL an empirical determination of the diffuse galactic light is needed. In the dark cloud method any zodiacal light foreground, even an isotropic one, is eliminated. The scattered light from the dark cloud itself and the general diffuse galactic light from the surroundings enter with full weight but it can be empirically determined using the characteristic features in its spectrum. A small opaque dark cloud in the Galactic halo or – even better – in the outskirts of an external galaxy or in the intergalactic space could serve as a perfect 'EBL zero reference position'. Park and Kim [175] suggest that they have already seen a few such dark objects of sub-arcsecond size in the Hubble Deep Field. Besides similar optical searches also high resolution (sub)millimeter surveys could reveal such objects via their long-wavelength thermal radiation.

intra-cluster and intra-group stellar population. Their contribution to the stellar mass, their age, and their dynamics. A&A, 516, A41